\documentclass[journal]{IEEEtran}

\usepackage[pdftex, pdfborder={0 0 0}, colorlinks=false]{hyperref}
\hypersetup{pdfborder={0 0 0}, colorlinks = true, linkcolor=black, citecolor=black, filecolor=black, urlcolor=black}
\usepackage[all]{hypcap}
\usepackage{graphicx}
\usepackage{subfigure}
\usepackage{algorithm}
\usepackage{algorithmic}
\usepackage{enumerate}
\usepackage{soul}
\usepackage{etoolbox}
\usepackage{times,amsmath,epsfig}
\usepackage{amsfonts} 
\usepackage{float}
\usepackage{url}
\usepackage{tikz}
\usepackage{varwidth}
\usepackage{verbatim}
\usetikzlibrary{shapes,decorations,shadows}
\usetikzlibrary{decorations.pathmorphing}
\usetikzlibrary{decorations.shapes}
\usetikzlibrary{fadings}
\usetikzlibrary{patterns}
\usetikzlibrary{calc}
\usetikzlibrary{decorations.text}
\usetikzlibrary{decorations.footprints}
\usetikzlibrary{decorations.fractals}
\usetikzlibrary{shapes.gates.logic.IEC}
\usetikzlibrary{shapes.gates.logic.US}
\usetikzlibrary{fit,chains}
\usetikzlibrary{positioning}
\usepgflibrary{shapes}
\usetikzlibrary{scopes}

\newcommand{\bfX}{\mathbf{X}}   
\newcommand{\bfY}{\mathbf{Y}}   

\newcommand{\bfT}{\mathbf{T}}   

\newcommand{\calX}{\mathcal{X}} 
\newcommand{\calY}{\mathcal{Y}}
\newcommand{\calO}{\mathcal{O}}
\newcommand{\calK}{\mathcal{K}}

\newcommand{\calS}{\mathcal{S}}

\newcommand{\td}{\widetilde}
\newcommand{\what}{\widehat}

\newtheorem{theorem}{Theorem}[section]
\newtheorem{definition}[theorem]{Definition}
\newtheorem{lemma}[theorem]{Lemma}
\newtheorem{corollary}[theorem]{Corollary}

\newtheorem{example}[theorem]{Example}

\floatstyle{ruled}
\newfloat{algorithm}{htbp}{loa}
\floatname{algorithm}{Algorithm}

\graphicspath{{images/}}

\begin{document}

\title{On the Benefits of Sampling in Privacy Preserving Statistical Analysis on Distributed Databases}

\author{Bing-Rong~Lin,
        Ye~Wang,
        and~Shantanu~Rane
\thanks{B.~Lin is with the Department of Computer Science and Engineering, Pennsylvania State University, University Park, PA, email:{\tt blin@cse.psu.edu}.}
\thanks{Y.~Wang and S.~Rane are with Mitsubishi Electric Research Laboratories, Cambridge, MA, email: \{{\tt yewang},{\tt rane}\}{\tt @merl.com}.}}

\maketitle

\begin{abstract}
We consider a problem where mutually untrusting curators possess portions of a vertically partitioned database containing information about a set of individuals.
The goal is to enable an authorized party to obtain aggregate (statistical) information from the database while protecting the privacy of the individuals, which we formalize using Differential Privacy.
This process can be facilitated by an untrusted server that provides storage and processing services but should not learn anything about the database.
This work describes a data release mechanism that employs Post Randomization (PRAM), encryption and random sampling to maintain privacy, while allowing the authorized party to conduct an accurate statistical analysis of the data.
Encryption ensures that the storage server obtains no information about the database, while PRAM and sampling ensures individual privacy is maintained against the authorized party.
We characterize how much the composition of random sampling with PRAM increases the differential privacy of system compared to using PRAM alone.
We also analyze the statistical utility of our system, by bounding the estimation error --- the expected $\ell_2$-norm error
between the true empirical distribution and the estimated distribution --- as a function of the number of samples, PRAM noise, 
and other system parameters. Our analysis shows a tradeoff between increasing PRAM noise versus decreasing the number of samples to maintain a desired level of privacy, and we determine the optimal number of samples that balances this tradeoff and maximizes the utility.
In experimental simulations with the UCI ``Adult Data Set'' and with synthetically generated data, we confirm that the theoretically predicted optimal number of samples indeed achieves close to the minimal empirical error, and that our analytical error bounds match well with the empirical results.
\end{abstract}

\section{Introduction}

One of the most visible technological trends is the emergence and proliferation of large-scale data collection.
Public and private enterprises are collecting tremendous volumes of data on individuals, their activities, their preferences, their locations, their medical histories, and so on.
These enterprises include government organizations, healthcare providers, financial institutions, internet search engines, social networks, cloud service providers, and many other kinds of private companies.
Naturally, interested parties could potentially discern meaningful patterns and gain valuable insights if they were able to access and correlate the information across these large, distributed databases.
For example, a social scientist may want to determine the correlations between individual income with personal characteristics such as gender, race, age, education, etc., or a medical researcher may want to study the relationships between disease prevalence and individual environmental factors.
In such applications, it is imperative to maintain the privacy of individuals, while ensuring that the useful aggregate (statistical) information is only revealed to the authorized parties.
Indeed, unless the public is satisfied that their privacy is being preserved, they would not provide their consent for the collection and use of their personal information.
Additionally, the inherent distribution of this data across multiple curators present a significant challenge, as privacy concerns and policy would likely prevent these curators from directly sharing their data to facilitate statistical analysis in a centralized fashion.
Thus, tools must be developed for conducting statistical analysis on large and distributed databases, while addressing these privacy and policy concerns.

\begin{figure}
\centering
\includegraphics[width=3.4in]{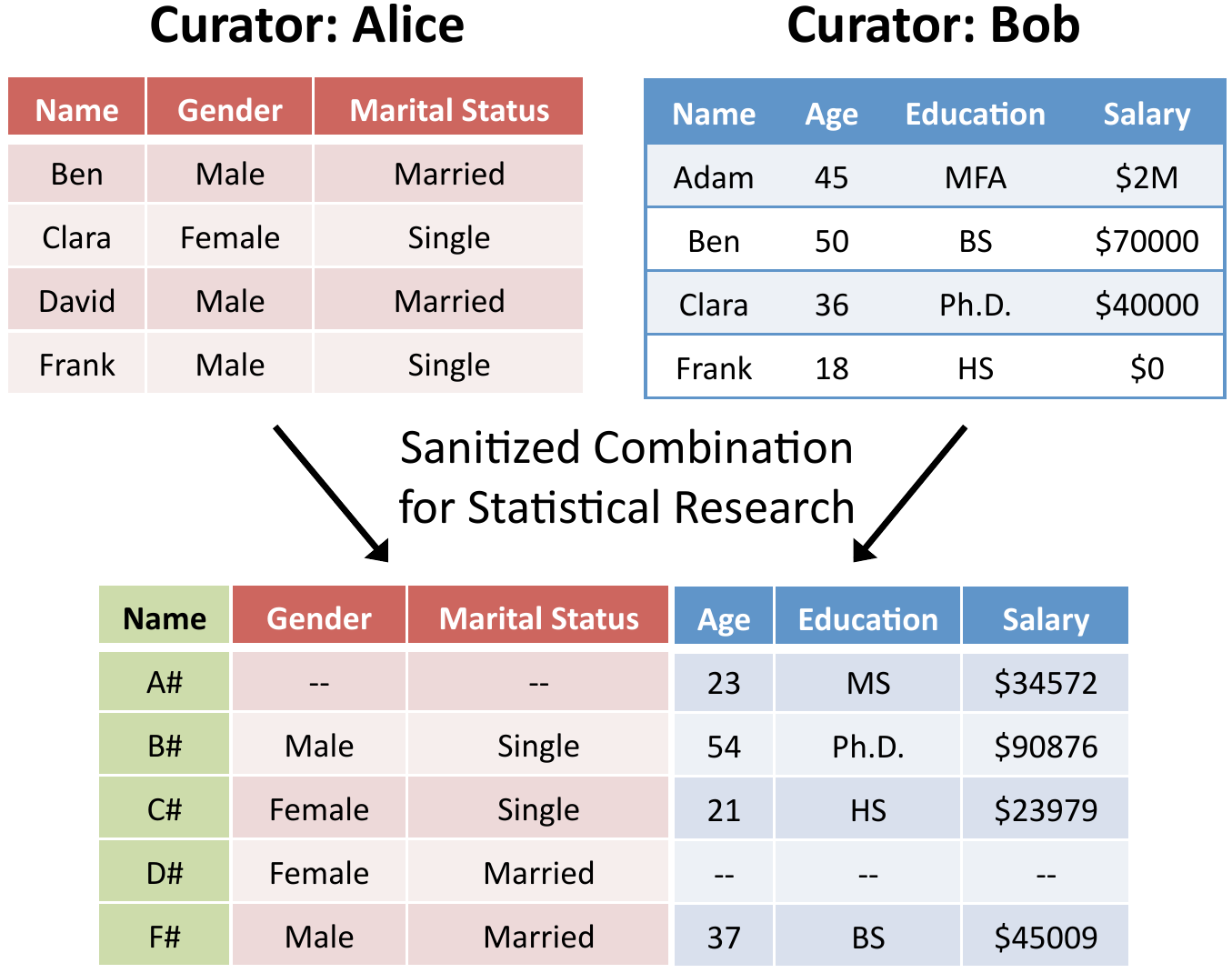}
\caption{An example in which curators Alice and Bob hold vertically partitioned data,
and a sanitized combination of their databases is made available for statistical
analysis.}
\label{fig:example-census}
\end{figure}

As an example, consider two curators Alice and Bob, who possess two databases
containing census-type information about individuals in a population, as shown in 
Figure~\ref{fig:example-census}.
Suppose that this data is to be combined and made available to authorized researchers studying salaries in the population, while ensuring that the privacy of the individual respondents is maintained.
Conceptually, a data release mechanism involves the ``sanitization'' of the data (via some form of perturbation or transform) to preserve individual privacy, before making it available for data analysis.
The suitability of the method used to sanitize the data is determined by the extent to
which rigorously defined privacy constraints are met.

Recent research has shown that conventional mechanisms for privacy, such as 
$k$-anonymization~\cite{samarati01microdata, samaratiS98:protecting} 
do not provide adequate privacy. Specifically, an informed adversary can link an arbitrary 
amount of side information to the anonymized database, and defeat the anonymization 
mechanism~\cite{narayanan09ssp}. In response to vulnerabilities of simple
anonymization mechanisms, a stricter notion of privacy ---  Differential 
Privacy~\cite{dwork08tamc,dwork09jpc} --- has been developed in recent
years. Informally, differential privacy ensures that the result of a function computed 
on a database of respondents is almost insensitive to the presence or absence of 
a particular respondent. A more formal way of stating this is that when the 
function is evaluated on adjacent databases (differing in only one respondent), 
the probability of outputting the same result is almost unchanged.
 
Mechanisms that provide differential privacy typically involve \emph{output} perturbation,
e.g., when Laplacian noise is added to the result of a
function computed on a database, it provides differential privacy to the individual
respondents in the database~\cite{smithlearn, diffrERM}.  
Nevertheless, it can be shown that \emph{input} perturbation approaches such as the 
randomized response mechanism~\cite{warner65asa, warner71asa} -- where
noise is added to the data itself -- also provide differential privacy to the respondents.
In this work, we are interested in a privacy mechanism that achieves three goals. Firstly,
the mechanism protects the privacy of individual respondents in a database. We achieve this through a privacy mechanism involving sampling and Post Randomization (PRAM)~\cite{gouweleeuwKWW98:PRAM}, which is a generalization of randomized response.
Secondly, the mechanism prevents unauthorized parties from learning anything about the data. We achieve this using random pads which can only be reversed by the authorized parties.
Thirdly, the mechanism achieves a superior tradeoff between privacy and utility 
compared to simply performing PRAM on the database. We show
that sampling the database enhances privacy with respect to 
the individual respondents while retaining the utility provided to an authorized researcher 
interested in the joint and marginal empirical probability distributions.

The idea of enhancing differential privacy via sampling, to the best of our knowledge, first
appeared in~\cite{smithlearn,adamsamplingdp} and was further developed by~\cite{NinghuiSamplingDP}. 
Theorem~\ref{thm:privacy} that we develop and prove herein is analogous to the
privacy amplification result of Theorem 1 in~\cite{NinghuiSamplingDP}, however, the theorems are
proved differently. Specifically, our proof requires an extra and non-trivial step because of 
the fact that the definition of differential privacy and sampling method in our setting are different.
In the definition of differential privacy used in~\cite{smithlearn, adamsamplingdp, NinghuiSamplingDP}, 
neighboring or adjacent databases are obtained by adding \emph{or} deleting an entry from the database
under consideration. This notion of adjacency cannot be used in our setting owing the fact that our setting 
involves perturbing the input data directly using techniques such as PRAM. In our work, 
an adjacent or neighboring database is obtained by replacing (i.e. deleting \emph{and} adding) a
single entry to the database under consideration. Further, the work in~\cite{smithlearn, 
adamsamplingdp, NinghuiSamplingDP} uses sampling with a fixed probability of including or 
excluding a sample, while our sampling mechanism is slightly different: the number of samples is 
fixed, and then sampling is carried out uniformly and without replacement based on the ratio 
of the number of samples to the size of the original database. This requires a different proof technique
that considers sets of possible samplings.

The more significant difference with respect to recent work is that, unlike~\cite{NinghuiSamplingDP}, 
we conduct a utility analysis, and derive a bound on 
the accuracy with which the desired statistical measures can be estimated, as a function of the
noise inserted for privacy and the number of samples. Our analysis reveals a privacy-utility
tradeoff between increasing PRAM noise versus decreasing the number of samples to maintain a desired 
level of differential privacy, and we determine the optimal number of samples that balances this 
tradeoff and maximizes the utility. We carry out experiments on both real-world and synthetically
generated data which confirm the existence of this tradeoff, and reveal that the experimentally
obtained optimal number of samples is very close to the number predicted by our analysis.

Another related work examines 
the effect of sampling on crowd-blending privacy~\cite{gehrkecrowd}. This is a strictly relaxed version of 
differential privacy, but it is shown that a pre-sampling step applied to a crowd-blending privacy mechanism
can achieve a desired amount of differential privacy. The scenario in our work differs from the 
treatment in~\cite{gehrkecrowd} in that we consider vertically partitioned distributed databases 
which are held by mutually untrusting curators. In our setting, computing joint statistics requires a join 
operation on the databases, which implies that individual curators cannot 
independently blend their 
respondents without altering the joint statistics across all databases.

The remainder of this paper is organized as follows: Section~\ref{sec:problemsetting} 
describes the multiparty problem setting, fixes notation and gives the privacy and utility definitions used in our analysis.
Section~\ref{sec:system} contains our main development, and begins by describing the 
mechanism itself, consisting of encryption via
random padding, randomized sampling, and data perturbation. It is shown that
sampling enhances the privacy of the individual respondents. An expression is
derived for the utility function, namely the expected $\ell_2$-norm error in the estimate of the 
joint distribution, in terms of the number of samples and the amount of noise introduced
by PRAM.
More importantly, the analysis reveals a tradeoff between the
number of samples and the perturbation noise. We conclude the section by deriving an 
expression for the optimal number of samples needed to maximize the utility function while 
achieving a desired level of privacy. In Section~\ref{sec:simulations}, the claims made in the theoretical analysis 
are tested experimentally with the UCI ``Adult Data Set''~\cite{Frank+Asuncion:2010} and with synthetically generated data.
In particular, the theoretically predicted optimal number of samples, that minimizes the error in the joint distribution, is found to agree closely with the experimental results.
Finally, Section~\ref{sec:discussion} summarizes the main results and concludes the paper with a discussion
on the practical considerations involved in performing privacy-preserving statistical analysis
using a combination of encryption, sampling and data perturbation.

\section{Problem Formulation}
\label{sec:problemsetting}

\begin{figure}\label{fig:setting-blockdiagram}
\begin{tikzpicture}
\tikzstyle{every node}=[font=\scriptsize ]
\node (bfX) at (-0.3,1) {\includegraphics[width=1cm]{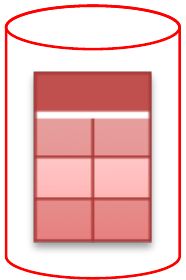}};
\node (bfY) at (-0.3,-1) {\includegraphics[width=1cm]{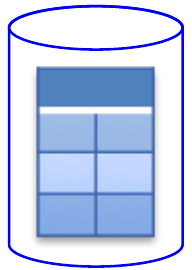}};
\node at (-0.3,0.1) {Alice};
\node at (-0.3,-1.9) {Bob};
\node (cloud) at (3.7,0) {\includegraphics[width=2.7cm]{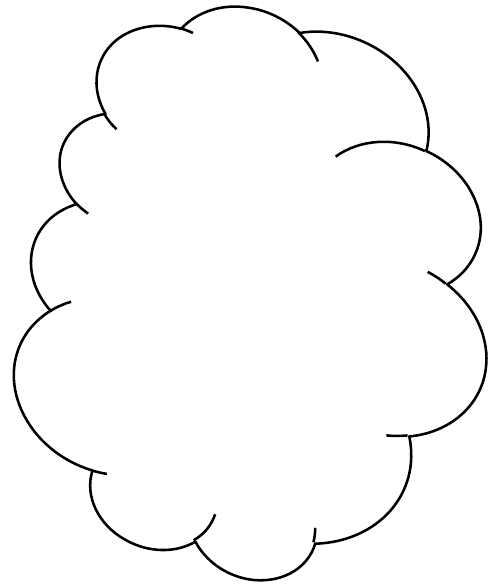}};
\node (server) at (3.5,-0.7) {\includegraphics[width=0.7cm]{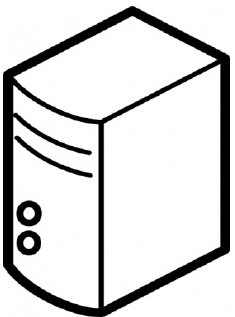}};
\node at (4.2,-0.7) {Cloud };
\node at (4.2,-0.9) { Server};
\node (researcher) at (6.0,0) {\includegraphics[width=1cm]{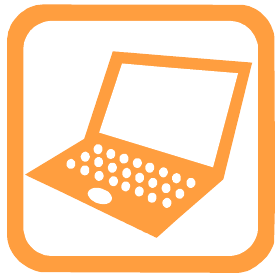}};
\node (rw) at (6.0,-0.6) {Researcher};
\node (x) at (0.4,1) { $X^n$  };
\node (y) at (0.4,-1) { $Y^n$  };
\node (dpx) at (2.2,1) { $O_A$  };
\node (dpy) at (2.2,-1) { $O_B$  };
\draw [->, line width=0.7] (x) -- (dpx) node [sloped,midway,above, minimum width=1cm] {$Enc(X^n ; K_A)$};
\draw [->, line width=0.7] (y) -- (dpy) node [sloped,midway,below, minimum width=1cm] {$Enc(Y^n ; K_B)$};
\draw [->, line width=0.7] (2.4,0.8) -- (3.1,0.2) node [sloped,midway,above, minimum width=1cm] {};
\draw [->, line width=0.7] (2.4,-0.8) -- (3.1,-0.2) node [sloped,midway,above, minimum width=1cm] {};
\node (ptype) at (3.9,0.0) { $M(O_A, O_B)$  };
\draw [->, line width=0.7] (ptype) -- (researcher) node [sloped,midway,above, minimum width=1cm] {};
\draw [-,densely dashed, line width=0.7] (1.2,1.5) -- (1.2,2) node [sloped,midway,above, minimum width=1cm] {};
\draw [-,densely dashed, line width=0.7] (1.2,2) -- (6.0,2) node [sloped,midway,below, minimum width=1cm] {Decryption key $K_A$};
\draw [->,densely dashed, line width=0.7] (6.0,2) -- (researcher) node [sloped,midway,above, minimum width=1cm] {};
\draw [-,densely dashed, line width=0.7] (1.2,-1.5) -- (1.2,-2) node [sloped,midway,above, minimum width=1cm] {};
\draw [-,densely dashed, line width=0.7] (1.2,-2) -- (6.0,-2) node [sloped,midway,above, minimum width=1cm] {Decryption key $K_B$};
\draw [->,densely dashed, line width=0.7] (6.0,-2) -- (rw) node [sloped,midway,above, minimum width=1cm] {};
\node (etw) at (7.5,0.3) {Estimate of};
\node (ejtype) at (7.6,0) { $\bfT_{\bfX, \bfY}$  };
\draw [->, line width=0.7] (researcher) -- (ejtype) node [sloped,midway,above, minimum width=1cm] {};
\end{tikzpicture}
\caption{Curators Alice and Bob independently encrypt their databases and provide it to a cloud server. The cloud server will sanitize the joint data. A researcher with decryption key can derive joint statistics or joint type based on the sanitized data, without compromising the privacy of individual database respondents. Neither the statistics nor the individual data entries are revealed to the cloud server.}
\end{figure}
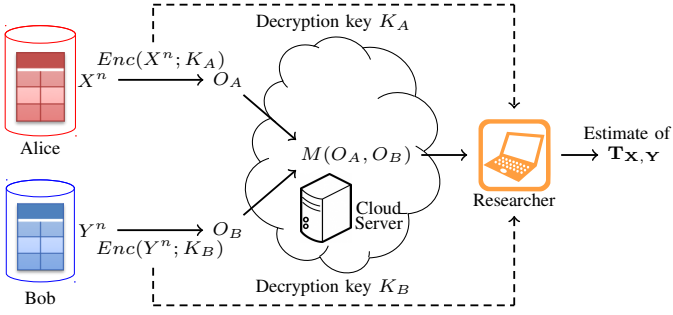

In this section, we present our general problem setup, wherein database curators wish to release data to enable privacy-preserving data analysis by an authorized party.
For ease of exposition, we present our problem formulation and results with two data curators, Alice and Bob, however our methods can easily be generalized to more than two curators.
Consider a data mining application in which 
Alice and Bob are mutually untrusting data curators, as shown in 
Figure~\ref{fig:setting-blockdiagram}. The two databases are to be made 
available for research with authorization granted by the data curators, such that statistical 
measures can be computed either on the individual databases, or on some combination of 
the two databases. Data curators should have flexible access control over the data. For example, 
if a researcher is granted access by Alice but not by Bob, then he/she can only access Alice's data. 
In addition, the cloud server should only host the data and not be able access the information. 
The data should be sanitized, before being released, to protect individual privacy. Altogether, we 
have the following privacy and utility requirements:
\begin{enumerate}
\item \textbf{Database Security}: Only researchers authorized by the data curators should be
able to extract statistical information from the database.
\item \textbf{Respondent Privacy}: Individual privacy of the respondents must be maintained 
against the cloud server as well as the researchers.
\item \textbf{Statistical Utility}: An \emph{authorized} researcher, i.e., one possessing appropriate
keys, should be able to compute the 
joint and marginal distributions of the data provided by Alice and Bob. 
\item \textbf{Complexity}:  The overall communication and computation requirements of the 
system should be reasonable.
\end{enumerate}

In the following sections, we will present our system framework and formalize the notions
of privacy and utility.

\subsection{Type and Matrix Notation}

The {\em type} (or empirical distribution) of a sequence $X^n$ is defined as the mapping $T_{X^n}: \mathcal{X} \rightarrow [0,1]$ given by
\[
\forall x \in \calX, \quad T_{X^n} (x) := \frac{|\{i : X_i = x\}|}{n}.
\]
The {\em joint type} of two sequences $X^n$ and $Y^n$ is defined as the mapping $T_{X^n, Y^n} : \calX \times \calY \rightarrow [0,1]$ given by
\[
\forall (x,y) \in \calX \times \calY, \quad T_{X^n, Y^n}(x,y) := \frac{|\{i : (X_i,Y_i) = (x,y)\}|}{n}.
\]

For notational convenience, when working with finite-domain type/distribution functions, we will drop the arguments to represent and use these functions as vectors/matrices.
For example, we can represent a distribution function $P_X : \calX \rightarrow [0,1]$ as the $|\calX| \times 1$ column-vector $P_X$, with its values arranged according to a fixed consistent ordering of $\calX$.
Thus, with a slight abuse of notation, using the values of $\calX$ to index the vector, the ``$x$''-th element of the vector, $P_X[x]$, is given by $P_X(x)$.
Similarly, a conditional distribution function $P_{Y|X}: \calY \times \calX \rightarrow [0,1]$ can be represented as a $|\calY| \times |\calX|$ matrix $P_{Y|X}$, defined by $P_{Y|X}[y,x] := P_{Y|X}(y|x)$.
For example, by utilizing this notation, the elementary probability identity
\[
\forall y \in \calY, \quad P_Y(y) = \sum_{x \in \calX} P_{Y|X}(y|x) P_X(x),
\]
can be written in matrix form as simply $P_Y = P_{Y|X}P_X$.

\subsection{System Framework}

{\bf Database Model:}
The data table held by Alice is modeled as a sequence $X^n := (X_1, X_2, \ldots, X_n)$, with each $X_i$ taking values in the finite-alphabet $\calX$.
Likewise, Bob's data table is modeled as a sequence of random variables $Y^n := (Y_1, Y_2, \ldots, Y_n)$, with each $Y_i$ taking values in the finite-alphabet $\calY$.
The length of the sequences, $n$, represents the total number of respondents
in the database, and each $(X_i, Y_i)$ pair represents the data of the respondent
$i$ collectively held by Alice and Bob, with the alphabet $\calX \times \calY$ representing the domain of each respondent's data.

{\bf Data Processing and Release:}
The curators each apply a data release mechanism to their respective data tables to produce an encryption of their data for the cloud server and a decryption key to be relayed to the researcher. These mechanisms are denoted by the randomized mappings $F_A : \calX^n \rightarrow \calO_A \times \calK_A$ and $F_B : \calY^n \rightarrow \calO_B \times \calK_B$, where $\calK_A$ and $\calK_B$ are suitable key spaces, and $\calO_B$ and $\calO_A$ are suitable encryption spaces.
The encryptions and keys are produced and given by
\begin{eqnarray*}
(O_A, K_A) := F_A(X^n), \\
(O_B, K_B) := F_B(Y^n).
\end{eqnarray*}
The encryptions $O_A$ and $O_B$ are sent to the cloud server, which performs processing, and the keys $K_A$ and $K_B$ are later sent to the researcher.
The cloud server processes $O_A$ and $O_B$, producing an output $O$ via a random mapping $M : \calO_A \times \calO_B \rightarrow \calO$, as given by
\[
O := M(O_A,O_B).
\]

{\bf Statistical Recovery:}
To enable the recovery of the statistics of the database, the processed output $O$ is provided to the researcher via the cloud server, and the encryption keys $K_A$ and $K_B$ are provided by the curators.
The researcher produces an estimate of the joint type (empirical distribution) of Alice and Bob's sequences, denoted by $\what T_{X^n,Y^n}$, as a function of $O$, $K_A$, and $K_B$, as given by
\[
\what T_{X^n,Y^n} := g(O, K_A, K_B),
\]
where $g : \calO \times \calK_A \times \calK_B \rightarrow [0,1]^{\calX \times \calY}$ is the reconstruction function.

The objective is to design a system within the above framework, by specifying the mappings $F_A$, $F_B$, $M$, and $g$, that optimize the system performance requirements, which are formulated in the next subsection.

\subsection{Privacy and Utility Conditions}
\label{sec:formulation}

In this subsection, we formulate the privacy and utility requirements for our problem framework.

{\bf Privacy against the Server:}
In the course of system operation, the data curators do not want reveal any information about their data tables (not even aggregate statistics) to the cloud server.
A strong statistical condition that guarantees this security is the requirement of statistical independence between the data tables and the encrypted versions held by the server.
The statistical requirement of independence guarantees security even against an adversarial server with unbounded resources, and does not require any unproved assumptions.

{\bf Respondent Privacy:}
The data pertaining to a respondent should be kept private from all other parties, including
any authorized researchers who aim to recover the statistics. We formalize this notion using
$\epsilon$-differential privacy for the respondents as follows:

\begin{definition}\cite{diffprivacy}
Given the above framework, the system provides $\epsilon$-differential privacy if for all databases $(x^n, y^n)$ and $(\dot{x}^n, \dot{y}^n)$ in $\calX^n \times \calY^n$, within Hamming distance $d_H((x^n, y^n), (\dot{x}^n, \dot{y}^n)) \leq 1$, and all $\calS \subseteq \calO \times \calK_A \times \calK_B$,
\begin{align*}
& \Pr \big[ (O, K_A, K_B) \in \calS \big| (X^n, Y^n) = (x^n, y^n) \big] \\
&\quad \leq e^\epsilon \Pr \big[ (O, K_A, K_B) \in \calS \big| (X^n, Y^n) = (\dot{x}^n, \dot{y}^n) \big]
\end{align*}
\end{definition}

This rigorous definition of privacy is widely used and satisfies the privacy axioms of~\cite{privaxioms,privaxioms:journal}.
Under the assumption that the respondents' data is i.i.d., this definition results in a strong privacy guarantee: an attacker with knowledge of all except one of the respondents cannot recover the data of the sole missing respondent~\cite{Kifer11NoFreeLunch}.

{\bf Utility for Authorized Researchers:}
The utility of the estimate is measured by the expected $\ell_2$-norm error of this estimated type vector, given by
\begin{align*}
& E \big\| \what T_{X^n,Y^n} - T_{X^n,Y^n} \big\|_2 \\
& \quad := \sqrt{\sum_{(x,y) \in \calX \times \calY}
   \big| \what T_{X^n,Y^n}(x,y) - T_{X^n,Y^n}(x,y) \big|^2},
\end{align*}
with the goal being the minimization of this error.

{\bf System Complexity:}
The communication and computational complexity of the system are also of concern.
The computational complexity can be captured by the complexity of implementing the mappings ($F_A$, $F_B$, $M$ and $g$) that specify a given system.
Ideally, one should aim to minimize the computational complexity of all of these mappings, simplifying the operations that each party must perform.
The communication requirements is given by the cardinalities of the symbol alphabets ($\calO_A$, $\calO_B$, $\calK_A$, $\calK_B$, and $\calO$).
The logarithms of these alphabet sizes indicate the sufficient length for the messages that must be transmitted in this system.

\section{Proposed System and Analysis}
\label{sec:system}

In this section, we will present the details of our system, and analyze its privacy and utility performance.
First, in Section~\ref{sec:SystemDescription}, we will describe how our system utilizes sampling and additive encryption, enabling a cloud server to join and perturb encrypted data in order to facilitate the release of sanitized data to the researcher.
Next, in Section~\ref{sec:panalysis}, we analyze the privacy of our system and show that sampling enhances privacy, thereby reducing the amount of noise that must be injected during the perturbation step in order to obtain a desired level of privacy.
Finally, in Section~\ref{sec:UtilityAnalysis}, we analyze the accuracy of the joint type reconstruction, producing a bound on the utility as a function of the system parameters, viz., the noise added during perturbation, and the
sampling factor.

\subsection{System Architecture} \label{sec:SystemDescription}

\begin{figure}
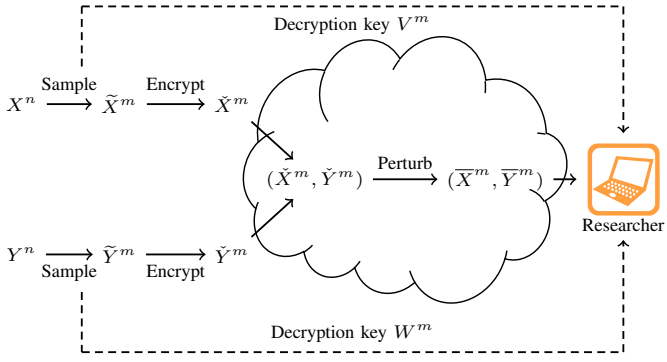
\label{fig:system:privacy}
\begin{tikzpicture}
\tikzstyle{every node}=[font=\scriptsize ]
\node [rotate=90] (cloud) at (5.5,0.1) {\includegraphics[width=3.7cm]{cloud}};
\node (researcher) at (8.4,0) {\includegraphics[width=1cm]{researcher2}};
\node (rw) at (8.4,-0.6) {Researcher};
\node (x) at (0.4,1) { $X^n$  };
\node (y) at (0.4,-1) { $Y^n$  };
\node (dpx) at (1.7,1) { $\td X^m $  };
\node (dpy) at (1.7,-1) { $\td Y^m $  };
\draw [->, line width=0.7] (x) -- (dpx) node [sloped,midway,above, minimum width=1cm] {Sample};
\draw [->, line width=0.7] (y) -- (dpy) node [sloped,midway,below, minimum width=1cm] {Sample};
\node (encx) at (3.2,1) { $\check X^m $  };
\node (ency) at (3.2,-1) { $\check Y^m $  };
\draw [->, line width=0.7] (dpx) -- (encx) node [sloped,midway,above, minimum width=1cm] {Encrypt};
\draw [->, line width=0.7] (dpy) -- (ency) node [sloped,midway,below, minimum width=1cm] {Encrypt};
\node (joinxy) at (4.3,0.0) { $(\check X^m, \check Y^m)$  };
\draw [->, line width=0.7] (encx) -- (joinxy) node [sloped,midway,above, minimum width=1cm] {};
\draw [->, line width=0.7] (ency) -- (joinxy) node [sloped,midway,above, minimum width=1cm] {};
\node (perturbxy) at (6.7,0.0) { $(\overline X^m, \overline Y^m)$  };
\draw [->, line width=0.7] (joinxy) -- (perturbxy) node [sloped,midway,above, minimum width=1cm] {Perturb};
\draw [->, line width=0.7] (perturbxy) -- (researcher) node [sloped,midway,above, minimum width=1cm] {};
\def\sideY{2.3}
\draw [-,densely dashed, line width=0.7] (1.2,1.5) -- (1.2,\sideY) node [sloped,midway,above, minimum width=1cm] {};
\draw [-,densely dashed, line width=0.7] (1.2,\sideY) -- (8.4,\sideY) node [sloped,midway,below, minimum width=1cm] {Decryption key $V^m$};
\draw [->,densely dashed, line width=0.7] (8.4,\sideY) -- (researcher) node [sloped,midway,above, minimum width=1cm] {};
\draw [-,densely dashed, line width=0.7] (1.2,-1.5) -- (1.2,-\sideY) node [sloped,midway,above, minimum width=1cm] {};
\draw [-,densely dashed, line width=0.7] (1.2,-\sideY) -- (8.4,-\sideY) node [sloped,midway,above, minimum width=1cm] {Decryption key $W^m$};
\draw [->,densely dashed, line width=0.7] (8.4,-\sideY) -- (rw) node [sloped,midway,above, minimum width=1cm] {};
\end{tikzpicture}
\caption{ Curators Alice and Bob independently encrypt their databases with a one time pad and provide it to a cloud 
server. The server samples $m$ respondents and then performs PRAM to guarantee privacy of the individual
database respondents. A researcher can derive joint statistics or joint type based on the sanitized data, without compromising the privacy of the respondents. Neither the statistics nor the individual data entries are revealed to the cloud server.}
\end{figure}

The data sanitization and release procedure is outlined by the following steps:
\begin{enumerate}
\item {\bf Sampling}: The curators randomly sample their data, producing shortened sequences.
\item {\bf Encryption}: The curators encrypt and send these shortened sequences to the cloud server.
\item {\bf Perturbation}: The cloud server combines and perturbs the encrypted sequences.
\item {\bf Release}: The researcher obtains the sanitized data from the server and the encryption keys from the curators, allowing the approximate recovery of data statistics.
\end{enumerate}
A key aspect of these steps is that the encryption and perturbation schemes are designed such that these operations
commute, thus allowing the server to essentially perform perturbation on the encrypted sequences, and for the 
authorized researcher to subsequently decrypt perturbed data.
In this section, we describe the details of each step from a theoretical perspective by applying mathematical abstractions and assumptions.
Later on, we will discuss practical implementations towards the realizing this system.
The overall data sanitization process is illustrated in Figure~\ref{fig:system:privacy}.

{\bf Sampling:}
The data curators reduce their length-$n$ database sequences $(X^n, Y^n)$ to $m$ randomly drawn samples.
We assume that these samples are drawn uniformly without replacement and that the curators will both sample at the same locations.
We will let $(\td X^m, \td Y^m) := (\td X_1, \ldots, \td X_m, \td Y_1, \ldots, \td Y_m)$ denote the intermediate result after sampling. 
Mathematically, the sampling result is described by, for all $i$ in $\{1, \ldots, m\}$,
\[
(\td X_i, \td Y_i) = (X_{I_i}, Y_{I_i}),
\]
where $I_1, \ldots, I_m$ are drawn uniformly without replacement from $\{1, \ldots, n\}$.

{\bf Encryption:} 
The data curators independently encrypt their sampled data sequences with an additive (one-time pad) encryption scheme.
To encrypt her data, Alice chooses an independent uniform key sequence $V^m \in \calX^m$, and produces the encrypted sequence
\[
\check X^m := \td X^m \oplus V^m := (\td X_1 + V_1, \ldots, \td X_m + V_m),
\]
where $\oplus$ denotes addition\footnote{The addition operation can be any suitably defined group addition operation over the finite set $\calX$.} applied element-by-element over the sequences.
Similarly, Bob encrypts his data, with the independent uniform key sequence $W^m \in \calY^m$, to produce the encrypted sequence
\[
\check Y^m := \td Y^m \oplus W^m := (\td Y_1 + W_1, \ldots, \td Y_m + W_m).
\]
Alice and Bob send these encrypted sequences to the cloud server, and will provide the keys to the researcher to enable data release.

{\bf Perturbation:}
The cloud server joins the encrypted data sequences, forming $((\check X_1,\check Y_1), \ldots, (\check X_m,
\check Y_m))$, and perturbs them by applying an independent PRAM mechanism, producing the perturbed results 
$(\overline X^m, \overline Y^m)$.
Each joined and encrypted sample, $(\check X_i,\check Y_i)$, is perturbed independently and identically according to a conditional distribution, $P_{\overline X, \overline Y | \check X, \check Y}$, that specifies a random mapping from $(\calX \times \calY)$ to $(\calX \times \calY)$.
Using the matrix $A := P_{\overline X, \overline Y | \check X, \check Y}$ to represent the conditional distribution, this operation can be described by
\[
P_{\overline X^m, \overline Y^m|\check X^m,\check Y^m}(\overline x^m,\overline x^m|\check x^m,\check x^m) = \prod_{i = 1}^m A[(\overline x_i,\overline y_i),(\check x_i,\check y_i)].
\]
By design, we specify that $A$ is a {\em $\gamma$-diagonal} matrix, for a parameter $\gamma > 1$, given by
\[
A[(\overline x,\overline y),(\check x,\check y)] :=
\begin{cases}
\gamma/q, & \text{if}\ (\overline x,\overline y) = (\check x,\check y), \\
1/q, & \text{o.w.},
\end{cases}
\]
where $q := (\gamma + |\calX||\calY| - 1)$ is a normalizing constant.

{\bf Release:}
In order to recover the data statistics, the researcher obtains the sampled, encrypted, and perturbed data sequences, $(\overline X^m, \overline Y^m)$, from the cloud server, and the encryption keys, $V^m$ and $W^m$, from the curators.
The researcher decrypts and recovers the sanitized data given by
\[
(\what X^m, \what Y^m) := (\overline X^m \oplus V^m, \overline Y^m \oplus W^m),
\]
which is effectively the data sanitized by sampling and PRAM (see Lemma~\ref{lem:Commutative} below).
The researcher produces the joint type estimate by inverting the matrix $A$ and multiplying it with the joint type of the sanitized data as follows
\[
\what T_{X^n,Y^n} := A^{-1} T_{\what X^m, \what Y^m}.
\]

Due to the $\gamma$-diagonal property of $A$, the PRAM perturbation is essentially an additive operation that commutes with the additive encryption.
This allows the server to perturb the encrypted data, with the perturbation being preserved when the encryption is removed.
The following Lemma summarizes this property, by stating that the decrypted, sanitized data recovered by the researcher, $(\what X^m, \what Y^m)$, is essentially the sampled data perturbed by PRAM.

\begin{lemma} \label{lem:Commutative}
Given the system described above, we have that
\[
P_{\what X^m, \what Y^m | \td X^m, \td Y^m} (\what x^m, \what y^m | \td x^m, \td y^m) = \prod_{i=1}^m A[(\what x_i, \what y_i),(\td x_i, \td y_i)].
\]
\end{lemma}

\subsection{Sampling Enhances Privacy} \label{sec:panalysis}
In this subsection, we will analyze the privacy of our proposed system.
Specifically, we show how sampling in conjunction with PRAM enhances the overall privacy for the respondents in comparison to using PRAM alone.
Note that if PRAM, with the $\gamma$-diagonal matrix $A$, was applied alone to the full databases, the resulting perturbed data would have $\ln(\gamma)$-differential privacy.
In the following lemma, we will show that the combination of sampling and PRAM results in sampled and perturbed data with enhanced privacy.

\begin{theorem} \label{thm:privacy}
The proposed system provides $\epsilon$-differential privacy for the respondents, where
\begin{equation} \label{eqn:privacy}
\epsilon = \ln \left( \frac{n+m(\gamma-1)}{n} \right).
\end{equation}
\end{theorem}

\IEEEproof
The researcher receives the perturbed and encrypted data from the server $O := (\overline X^m, \overline Y^m)$ and the keys $(K_A, K_B) := (V^m, W^m)$ from the curators.
However, since the sanitized data, $(\what X^m, \what Y^m)$, recovered by the researcher is a sufficient statistic for the original databases, that is, the following Markov chain holds
\[
(X^n, Y^n) - (\what X^m, \what Y^m) - (\overline X^m, \overline Y^m, V^m, W^m),
\]
we need only to show that, for all $(x^n, y^n)$, $(\dot{x}^n, \dot{y}^n)$, and $(\what x^m, \what y^m)$ in $\calX^n \times \calY^n$ with $d_H((x^n, y^n), (\dot{x}^n, \dot{y}^n)) = 1$, 
\[
\frac{P_{\what X^m, \what Y^m | X^n, Y^n}
        (\what x^m, \what y^m | x^n, y^n)}
     {P_{\what X^m, \what Y^m | X^n, Y^n}
        (\what x^m, \what y^m | \dot{x}^n, \dot{y}^n)} \leq e^{\epsilon},
\]
in order to prove $\epsilon$-differential privacy for the respondents.
Since $d_H((x^n, y^n), (\dot{x}^n, \dot{y}^n)) = 1$, the two database differ in only one location.
Let $k$ denote the location where $(x_k, y_k) \neq (\dot{x}_k, \dot{y}_k)$.

Before we proceed, we introduce some notation regarding sampling to facilitate the steps of our proof.
We will use the following notation for the set of all possible samplings
\[
\Theta := \big\{ \pi | \pi := (\pi_1, \ldots, \pi_m) \in \{1,\ldots,n\}^m, \pi_i \neq \pi_j, \forall i \neq j \big\}.
\]
The sampling locations $(I_1, \ldots, I_m)$ are uniformly drawn from the set $\Theta$.
We also define $\Theta_k := \{\pi \in \Theta\,|\,\, \exists\, i, \pi_i = k\}$ to denote the subset of samplings that select location $k$, and $\Theta_k^c := \Theta \setminus \Theta_k$ to denote the subset of samplings that do not select location $k$.
For $\pi \in \Theta_k$, we define $\Theta_k(\pi) := \{ \pi' \in \Theta_k^c | d_H(\pi,\pi') = 1\}$ as the subset of $\Theta_k^c$ that replaces the selection of location $k$ with any other non-selected location.
We will also slightly abuse notation by using $\pi \in \Theta$ as sampling function for the database sequences, that is, $\pi(X^n) := (X_{\pi_1}, \ldots, X_{\pi_m})$, and similarly for $\pi(Y^n)$.
Using the above notation, we can rewrite the following conditional probability,
\begin{align*}
& P_{\what X^m, \what Y^m | X^n, Y^n} (\what x^m, \what y^m | x^n, y^n) \\
& \  = \sum_{\pi \in \Theta} \frac{1}{|\Theta|}
    P_{\what X^m, \what Y^m | \td X^m, \td Y^m}
    \big( \what x^m, \what y^m | \pi(x^n), \pi(y^n) \big) \\
& \  = \frac{1}{|\Theta|} \Big[ \sum_{\pi \in \Theta_k}
    P_{\what X^m, \what Y^m | \td X^m, \td Y^m}
    \big( \what x^m, \what y^m | \pi(x^n), \pi(y^n) \big) \\
& \quad +  \sum_{\pi' \in \Theta_k^c}
    P_{\what X^m, \what Y^m | \td X^m, \td Y^m}
    \big( \what x^m, \what y^m | \pi'(x^n), \pi'(y^n) \big) \Big] \\
& \ = \frac{1}{|\Theta|} \Big[ \sum_{\pi \in \Theta_k} \Big(
    P_{\what X^m, \what Y^m | \td X^m, \td Y^m}
    \big( \what x^m, \what y^m | \pi(x^n), \pi(y^n) \big) \\
& \quad + \frac{1}{m} \sum_{\pi' \in \Theta_k(\pi)}
    P_{\what X^m, \what Y^m | \td X^m, \td Y^m}
    \big( \what x^m, \what y^m | \pi'(x^n), \pi'(y^n) \big) \Big) \Big],
\end{align*}
where in the last equality we have rearranged the summations to embed the summation over $\pi' \in \Theta_k^c$ into the summation over $\pi \in \Theta_k$.
Note that summing over all $\pi' \in \Theta_k(\pi)$ within a summation over all $\pi \in \Theta_k$ covers all $\pi' \in \Theta_k^c$, but overcounts each $\pi'$ exactly $m$ times since each $\pi' \in \Theta_k^c$ belongs to $m$ of the $\Theta_k(\pi)$ sets across all $\pi \in \Theta_k$. Hence, a $(1/m)$ term has been added to account for this overcount.

To ease the use of the above expansion, we introduce the following shorthand notation for the summation terms,
\begin{align*}
\alpha(\pi) & := P_{\what X^m, \what Y^m | \td X^m, \td Y^m}
    \big( \what x^m, \what y^m | \pi(x^n), \pi(y^n) \big) \\
\beta(\pi) & := P_{\what X^m, \what Y^m | \td X^m, \td Y^m}
    \big( \what x^m, \what y^m | \pi(\dot{x}^n), \pi(\dot{y}^n) \big).
\end{align*}
Thus, the following probability ratio can be written as
\begin{align*}
& \frac{P_{\what X^m, \what Y^m | X^n, Y^n}
        (\what x^m, \what y^m | x^n, y^n)}
       {P_{\what X^m, \what Y^m | X^n, Y^n}
        (\what x^m, \what y^m | \dot{x}^n, \dot{y}^n)} \\
& \quad = 
  \frac{\sum_{\pi \in \Theta_k} \big( \alpha(\pi) +
        \frac{1}{m} \sum_{\pi' \in \Theta_k(\pi)} \alpha(\pi') \big)}
       {\sum_{\pi \in \Theta_k} \big( \beta(\pi) +
        \frac{1}{m} \sum_{\pi' \in \Theta_k(\pi)} \beta(\pi') \big)} \\
& \quad \leq \max_{\pi \in \Theta_k}
  \frac{\alpha(\pi) + \frac{1}{m} \sum_{\pi' \in \Theta_k(\pi)} \alpha(\pi')}
       {\beta(\pi) + \frac{1}{m} \sum_{\pi' \in \Theta_k(\pi)} \beta(\pi')} \\
& \quad =
  \frac{\alpha(\pi^*) + \frac{1}{m} \sum_{\pi' \in \Theta_k(\pi^*)} \alpha(\pi')}
       {\beta(\pi^*) + \frac{1}{m} \sum_{\pi' \in \Theta_k(\pi^*)} \beta(\pi')},
\end{align*}
where $\pi^* \in \Theta_k$ denotes the sampling that maximizes the ratio.
Given the $\gamma$-diagonal structure of the matrix $A$, we have that
\[
\gamma^{-1} \alpha(\pi^*) \leq \beta(\pi^*),
\]
since $(\pi^*(x^n), \pi^*(y^n))$ and $(\pi^*(\dot{x}^n), \pi^*(\dot{y}^n))$ differ in only one location,
\[
\gamma^{-1} \alpha(\pi^*) \leq \alpha(\pi'), \quad \forall \pi' \in \Theta_k(\pi^*),
\]
since $(\pi^*(x^n), \pi^*(y^n))$ and $(\pi'(x^n), \pi'(y^n))$ differ in only one location, and 
\[
\alpha(\pi') = \beta(\pi'), \quad \forall \pi' \in \Theta_k(\pi^*),
\]
since $(\pi'(x^n), \pi'(y^n)) = (\pi'(\dot{x}^n), \pi'(\dot{y}^n))$.
Given these constraints, we can continue to bound the likelihood ratio as
\begin{align*}
& \frac{P_{\what X^m, \what Y^m | X^n, Y^n}
        (\what x^m, \what y^m | x^n, y^n)}
       {P_{\what X^m, \what Y^m | X^n, Y^n}
        (\what x^m, \what y^m | \dot{x}^n, \dot{y}^n)} \\
& \quad \leq
  \frac{\alpha(\pi^*) + \frac{1}{m} \sum_{\pi' \in \Theta_k(\pi^*)} \alpha(\pi')}
       {\beta(\pi^*) + \frac{1}{m} \sum_{\pi' \in \Theta_k(\pi^*)} \beta(\pi')} \\
& \quad \leq
  \frac{\alpha(\pi^*) + \frac{n-m}{m} \gamma^{-1} \alpha(\pi^*)}
       {\gamma^{-1} \alpha(\pi^*) + \frac{n-m}{m} \gamma^{-1} \alpha(\pi^*)} \\
& \quad = 
  \frac{n+m(\gamma-1)}{n} = e^{\epsilon},
\end{align*}
thus finishing the proof by bounding the likelihood ratio with $e^\epsilon$.
\IEEEQED

To show $\epsilon$-differential privacy for a given $\epsilon$, we only need to upperbound the probability ratio by $e^\epsilon$, as done in the above proof.
A natural question is if this bound is tight, that is, whether there exists a smaller $\epsilon$ for which the bound also holds, hence making the system more private.
With the following example, we show that the value for $\epsilon$ given in Theorem~\ref{thm:privacy} is tight.
\begin{example}
Let $a$ and $b$ be two distinct elements in $(\calX \times \calY)$.
Let $(x^n,y^n) = (b, a, a, \ldots, a)$, $(\dot{x}^n, \dot{y}^n) = (a, a, \ldots, a)$ and $(\what{x}^m, \what{y}^m) = (b, b, \ldots, b)$.
Let $E$ denote the event that the first element (where the two databases differ) is sampled, which occurs with probability $(m/n)$.
We can determine the likelihood ratio as follows
\begin{align*}
& \frac{P_{\what X^m, \what Y^m | X^n, Y^n}
        (\what x^m, \what y^m | x^n, y^n)}
       {P_{\what X^m, \what Y^m | X^n, Y^n}
        (\what x^m, \what y^m | \dot{x}^n, \dot{y}^n)} \\
& \quad =
  \frac{\Pr[E] \gamma(1/q)^m + (1-\Pr[E])(1/q)^m}
       {(1/q)^m} \\
& \quad = 
  \frac{n+m(\gamma-1)}{n} = e^{\epsilon}.
\end{align*}
Thus, the value of $\epsilon$ given by Theorem~\ref{thm:privacy} is tight.
\end{example}

As a consequence of the privacy analysis of Theorem~\ref{thm:privacy}, we have that for given system parameters of database length $n$, number of samples $m$, and desired level privacy $\epsilon$, the level of PRAM perturbation, specified by the $\gamma$ parameter of the matrix $A$, must be
\begin{equation} \label{eqn:gamma}
\gamma = 1 + \frac{n}{m}(e^\epsilon - 1).
\end{equation}

Privacy against the server is obtained as a consequence of the one-time-pad encryption performed on the data prior to transmission to the server.
It is straightforward to verify that the encryptions received by the server are statistically independent of the original database as a consequence of the independence and uniform randomness of the keys.

\subsection{Utility Analysis} \label{sec:UtilityAnalysis}
In this subsection, we will analyze the utility of our proposed system.
Our main result is a theoretical bound on the expected $\ell_2$-norm of the joint type estimation error.
Analysis of this bound will illustrate the tradeoffs between utility and privacy level $\epsilon$ as function of sampling parameter $m$ and PRAM perturbation level $\gamma$.
Given this error bound, we can compute the optimal sampling parameter $m$ for minimizing the error bound while achieving a fixed privacy level $\epsilon$.

\begin{theorem} \label{thm:utility}
For our proposed system, the expected $\ell_2$-norm of the joint type estimate is bounded by
\begin{equation} \label{eqn:utilitybound}
E \big\| \what T_{X^n,Y^n} - T_{X^n,Y^n} \big\|_2
  \leq \frac{c\sqrt{|\calX||\calY|} + 1}{\sqrt{m}}.
\end{equation}
where $c$ is the condition number of the $\gamma$-diagonal matrix $A$, given by
\[
c = 1 + \frac{|\calX||\calY|}{\gamma - 1}.
\]
\end{theorem}

\IEEEproof
The expected $\ell_2$-norm error is given by
\[
E \big\| \what T_{X^n,Y^n} - T_{X^n,Y^n} \big\|_2 = E \big\| A^{-1} T_{\what X^m, \what Y^m} - T_{X^n,Y^n} \big\|_2.
\]
Applying the triangle inequality, we can bound the error as the sum of the error introduced by sampling and the error introduced by PRAM, as follows,
\begin{align*}
& E \big\| A^{-1} T_{\what X^m, \what Y^m} - T_{X^n,Y^n} \big\|_2 \leq \\ 
& \quad E \big\| T_{\td X^m, \td Y^m} - T_{X^n,Y^n} \big\|_2
  + E \big\| A^{-1} T_{\what X^m, \what Y^m} - T_{\td X^m, \td Y^m} \big\|_2.
\end{align*}

We will analyze and bound the sampling error,
\[
E \big\| T_{\td X^m, \td Y^m} - T_{X^n,Y^n} \big\|_2,
\]
by utilizing the smoothing theorem by first bounding the conditional expectation
\[
E \left[ \big\| T_{\td X^m, \td Y^m} - T_{X^n,Y^n} \big\|_2 \Big| T_{X^n,Y^n} \right].
\]
For a given $(x,y) \in \calX \times \calY$, the sampled type, $T_{\td X^m, \td Y^m}(x,y)$, conditioned on $T_{X^n,Y^n}$, is a hypergeometric random variable normalized by $m$, with expectation and variance given by
\begin{align*}
& E \big[ T_{\td X^m, \td Y^m}(x,y) \big| T_{X^n,Y^n} \big] = T_{X^n,Y^n}(x,y), \\
& \mathrm{Var} \big[ T_{\td X^m, \td Y^m}(x,y) \big| T_{X^n,Y^n} \big] = \\
& \quad \frac{nT_{X^n,Y^n}(x,y) \big( n - nT_{X^n,Y^n}(x,y) \big) (n - m)}{mn^2(n-1)} \\
& \quad \leq \frac{T_{X^n,Y^n}(x,y)}{m}.
\end{align*}

Applying Jensen's inequality to the conditioned sampling error yields
\begin{align*}
& E \left[ \big\| T_{\td X^m, \td Y^m} - T_{X^n,Y^n} \big\|_2 \Big| T_{X^n,Y^n} \right] \\
& \quad \leq \sqrt{\sum_{(x,y) \in \calX \times \calY} \mathrm{Var} \big[ T_{\td X^m, \td Y^m}(x,y) \big| T_{X^n,Y^n} \big] } \leq \frac{1}{\sqrt{m}}.
\end{align*}

Applying the smoothing theorem, the sampling error can be bounded by
\begin{align*}
& E \big\| T_{\td X^m, \td Y^m} - T_{X^n,Y^n} \big\|_2 = \\
& \quad E_{T_{X^n,Y^n}} \left[ E \left[ \big\| T_{\td X^m, \td Y^m} - T_{X^n,Y^n} \big\|_2 \Big| T_{X^n,Y^n} \right] \right] \\
& \quad \leq \frac{1}{\sqrt{m}}.
\end{align*}

Next, to analyze and bound the PRAM error given by
\[
E \big\| A^{-1} T_{\what X^m, \what Y^m} - T_{\td X^m, \td Y^m} \big\|_2,
\]
we will make use of the following linear algebra lemma.

\begin{lemma}
Let $A$ be an invertible matrix and $(x, y)$ be vectors that satisfy $Ax = y$. For any vectors $(\hat{x}, \hat{y})$ such that $\hat{x} = A^{-1} \hat{y}$, we have
\[
\frac{\|\hat{x} - x\|}{\|x\|} \leq c \frac{\|\hat{y} - y\|}{\|y\|},
\]
where $c$ is the condition number of the matrix $A$.
\end{lemma}

To bound the PRAM error, we will make use of the following consequence of this lemma,
\begin{align*}
& \big\| A^{-1} T_{\what X^m, \what Y^m} - T_{\td X^m, \td Y^m} \big\|_2 \\ 
& \quad \leq c \frac{\| T_{\td X^m, \td Y^m} \|_2}{\| A T_{\td X^m, \td Y^m} \|_2} \big\| T_{\what X^m, \what Y^m} - A T_{\td X^m, \td Y^m} \big\|_2,
\end{align*}
which allows us to bound the conditional expectation of the PRAM error as follows,
\begin{align*}
& E \left[ \big\| A^{-1} T_{\what X^m, \what Y^m} - T_{\td X^m, \td Y^m} \big\|_2
  \Big| T_{\td X^m, \td Y^m} \right] \\
& \quad \leq c \frac{\| T_{\td X^m, \td Y^m} \|_2}{\| A T_{\td X^m, \td Y^m} \|_2} 
  E \left[ \big\| T_{\what X^m, \what Y^m} - A T_{\td X^m, \td Y^m} \big\|_2
  \Big| T_{\td X^m, \td Y^m} \right].
\end{align*}

For a given $(x,y) \in \calX \times \calY$, the perturbed and sampled type, $T_{\what X^m, \what Y^m}(x,y)$, conditioned on $T_{\td X^m, \td Y^m}$, is a poisson-binomial random variable normalized by $m$ with expectation and variance given by
\begin{align*}
& E \big[ T_{\what X^m, \what Y^m}(x,y) \big| T_{\td X^m, \td Y^m} \big]
  = (AT_{\td X^m, \td Y^m})[x,y], \\
& \mathrm{Var} \big[ T_{\what X^m, \what Y^m}(x,y)
  \big| T_{\td X^m, \td Y^m} \big] = \\
& \quad \frac{1}{m} \hspace{-12pt} \sum_{(x',y') \in \calX \times \calY} \hspace{-12pt} T_{\td X^m, \td Y^m} (x',y') A[(x,y),(x',y')] (1-A[(x,y),(x',y')]).
\end{align*}

We can bound the following conditional expectation using Jensen's inequality to yield
\begin{align*}
& E \left[ \big\| T_{\what X^m, \what Y^m} - A T_{\td X^m, \td Y^m} \big\|_2 
  \Big| T_{\td X^m, \td Y^m} \right] \\
& \quad \leq \sqrt{ \sum_{(x,y) \in \calX \times \calY}
  \mathrm{Var} \left[ T_{\what X^m, \what Y^m} (x,y)
  \Big| T_{\td X^m, \td Y^m} \right] } \\
& \quad = \Bigg[ \frac{1}{m} \sum_{(x,y) \in \calX \times \calY}
  \sum_{(x',y') \in \calX \times \calY}
  T_{\td X^m, \td Y^m} (x',y') \\ 
& \quad \quad \quad A[(x,y),(x',y')] \big(1-A[(x,y),(x',y')]\big) \Bigg]^{1/2} \\
& \quad \leq \Bigg[ \frac{1}{m} \sum_{(x',y') \in \calX \times \calY} \hspace{-10pt}
  T_{\td X^m, \td Y^m} (x',y') \hspace{-10pt} \sum_{(x,y) \in \calX \times \calY} \hspace{-10pt}
  A[(x,y),(x',y')] \Bigg]^{1/2} \\
& \quad = \frac{1}{\sqrt{m}}.
\end{align*}

Combining equations yields the bound
\begin{align}
& E \left[ \big\| A^{-1} T_{\what X^m, \what Y^m} - T_{\td X^m, \td Y^m} \big\|_2
  \Big| T_{\td X^m, \td Y^m} \right] \nonumber \\
& \quad \leq \frac{c}{\sqrt{m}} \frac{\| T_{\td X^m, \td Y^m} \|_2}{\| A T_{\td X^m, \td Y^m} \|_2} \nonumber \\
& \quad \leq \frac{c}{\sqrt{m}} \frac{\sqrt{|\calX||\calY|} \| T_{\td X^m, \td Y^m} \|_1}{\| A T_{\td X^m, \td Y^m} \|_1} \label{eqn:NormBound}\\
& \quad = c \sqrt{\frac{|\calX||\calY|}{m}}, \nonumber
\end{align}
which, upon applying the smoothing theorem, yields the following bound on the PRAM error
\begin{align*}
& E \left[ \big\| A^{-1} T_{\what X^m, \what Y^m} - T_{\td X^m, \td Y^m} \big\|_2 \right] \\
& \quad = E_{T_{\td X^m, \td Y^m}} \left[ 
  E \left[ \big\| A^{-1} T_{\what X^m, \what Y^m} - T_{\td X^m, \td Y^m} \big\|_2
  \Big| T_{\td X^m, \td Y^m} \right] \right] \\
& \quad \leq c \sqrt{\frac{|\calX||\calY|}{m}}.
\end{align*}

Combining the individual bounds on the sampling and PRAM error via the triangle inequality yields the following bound on expected norm-2 error of the type estimate formed from the sampled and perturbed data,
\[
E \big\| A^{-1} T_{\what X^m, \what Y^m} - T_{X^n,Y^n} \big\|_2
  \leq \frac{c\sqrt{|\calX||\calY|} + 1}{\sqrt{m}}.
\]

Since $A$ is a $\gamma$-diagonal matrix, its condition number $c$ is given by
\[
c = 1 + \frac{|\calX||\calY|}{\gamma - 1}.
\]
\IEEEQED

Given a fixed PRAM perturbation parameter $\gamma$, the error bound decays on the order of $O(1/\sqrt{m})$ as a function of the sampling parameter $m$.
However, as $m$ increases, $\epsilon$ as given in Equation~(\ref{eqn:privacy}) also grows, decreasing privacy.
However, when we fix the overall privacy level $\epsilon$, by adjusting $\gamma$ as a function of $m$, as given by Equation~(\ref{eqn:gamma}), in order to maintain the desired level of privacy, we observe that increasing $m$ too much will cause the error bound to expand.
Intuitively, this can be explained as by having $m$ too large, we need to increase the PRAM perturbation through lowering $\gamma$ to maintain the same level of privacy, which has the adverse effect of increasing the error bound through the condition number $c$.
On the other hand, by having $m$ too small, too few samples are taken resulting in an inaccurate type estimate.
This balance in adjusting the sampling parameter $m$ shows that there is an optimal sample size $m$ as a function of the desired level of privacy $\epsilon$ and other system parameters.
The theoretically optimal sample size $m$ for the error upper bound is given by the following corollary.

\begin{corollary}
The optimal sampling parameter $m^*$ that optimizes the error bound of Equation~(\ref{eqn:utilitybound}) is
\begin{equation} \label{eqn:optimalsampling}
m^* = \frac{ n \left(1 + \sqrt{|\calX||\calY|}\right)(e^{\epsilon}-1) }
{(|\calX||\calY|)^{\frac{3}{2}}}.
\end{equation}
\end{corollary}

\IEEEproof
By combining equations for the expected error bound, Equation~(\ref{eqn:utilitybound}), and the required level of $\gamma$, Equation~(\ref{eqn:gamma}), we have
\begin{align*}
\frac{c\sqrt{|\calX||\calY|} + 1}{\sqrt{m}} &= \frac{\left(1+\frac{|\calX||\calY|}{\gamma-1}\right)\sqrt{|\calX||\calY|} + 1}{\sqrt{m}} \\
&= \frac{1 + \sqrt{|\calX||\calY|}}{\sqrt{m}} + \frac{(|\calX||\calY|)^{\frac{3}{2}}}{\sqrt{m} (\gamma-1)} \\
&= \frac{1 + \sqrt{|\calX||\calY|}}{\sqrt{m}} + \frac{(|\calX||\calY|)^{\frac{3}{2}} \sqrt{m}}{n(e^\epsilon - 1)}.
\end{align*}
By setting the derivative of this expression to zero, we can solve to find the optimal $m$,
\begin{align}
& \left(1 + \sqrt{|\calX||\calY|}\right) \left(\frac{-m^{-\frac{3}{2}}}{2}\right) 
    + \frac{(|\calX||\calY|)^{\frac{3}{2}}}{n(e^\epsilon - 1)} \left(\frac{m^{-\frac{1}{2}}}{2}\right) = 0 \nonumber \\
& \quad \iff m^* = \frac{ n \left(1 + \sqrt{|\calX||\calY|}\right)(e^{\epsilon}-1) }{(|\calX||\calY|)^{\frac{3}{2}}}. \label{equ:optimalsamples}
\end{align}
\IEEEQED

\begin{figure}\label{fig:system:utility}
\begin{tikzpicture}
\tikzstyle{every node}=[font=\scriptsize ]
\node (researcher) at (7.0,0) {\includegraphics[width=1cm]{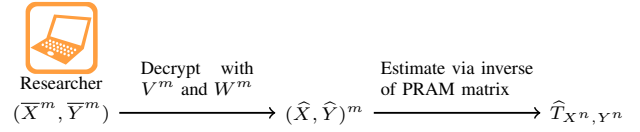}};
\node (rw) at (7.0,-0.6) {Researcher};
\node (encData) at (7,-1) { $ (\overline X^m, \overline Y^m) $  };
\node (pramData) at (10.5,-1) { $(\what X  , \what Y )^m $  };
\draw [->, line width=0.7] (encData) -- (pramData) node [sloped,midway,above=0.1cm, minimum width=1.4cm] { \begin{varwidth}{1.5cm} Decrypt with $V^m$ and $W^m$ \end{varwidth} };
\node (etype) at (14,-1) { $\what{T}_{X^n, Y^n} $  };
\draw [->, line width=0.7] (pramData) -- (etype) node [sloped,midway,above=0.1cm, minimum width=2.2cm] { \begin{varwidth}{2.2cm} Estimate via inverse of PRAM matrix \end{varwidth} };
\end{tikzpicture}
\caption{Authorized researchers apply decryption keys $V^m$ and $W^m$ from Alice and Bob to decrypt the message and obtain the perturbed samples. They then use the inverse of the PRAM matrix to estimate the true type.}
\end{figure}

\section{Experimental Results}
\label{sec:simulations}

In order to validate our theoretical results, we conducted experiments that simulated our proposed system using the UCI ``Adult Data Set''~\cite{Frank+Asuncion:2010} and synthetically generated data.
The UCI ``Adult Data Set'' was extracted from the 1994 Census database and consists of personal information for over 48 thousand individuals, with various attributes including age, education, marital status, occupation, gender, race, income, etc.

\begin{figure}
\centering
\includegraphics[width=3.2in]{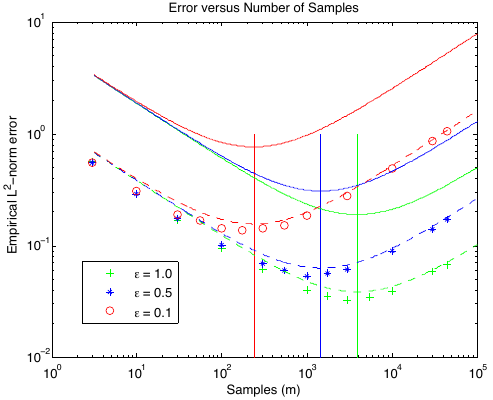}
\caption{The experimental results from simulations with the UCI dataset are plotted as points alongside the theoretical results. The experiments were conducted for three privacy levels ($\epsilon$) and across a range of number of samples ($m$). Each data point represents the expected $\ell_2$-norm of the type error estimated as the empirical mean over $1000$ independent experiments.
The solid curves illustrate the theoretical error bound, and the solid vertical lines illustrate theoretically optimal number of samples at each privacy level.
The dashed lines correspond to the error bound divided by a factor of $\sqrt{|\calX||\calY|}$ to illustrate that the bounds seem to capture the correct shape, albeit being loose by a multiplicative factor.}
\label{fig:UCI-Results}
\end{figure}

For the first set of experiments, we reduced the cardinality of the attribute set by considering only a subset of the 
attributes as well as quantizing some attributes into categories. Specifically, we used education 
(quantized to ``no college'', ``some college'', or ``post-graduate degree''), marital status (quantized 
to ``married'' or ``single/divorced/widowed''), gender (inherently categorized as ``male'' or ``female''), and 
salary (inherently categorized as ``over 50K'' or ``50K or less''), resulting in a total attribute set cardinality 
of $|\calX||\calY| = 24$. We also discarded any individuals where there was missing information in any of 
these attributes, reducing the size of the total dataset to 45222 individuals. In this and the remaining 
experiments, while varying the sampling parameter $m$ and overall privacy level $\epsilon$, we set the level of PRAM 
perturbation $\gamma$ as dictated by Equation~(\ref{eqn:gamma}). The results of the simulations with 
the UCI ``Adult Data Set'' are presented in Figure~\ref{fig:UCI-Results}.
The data points show the simulation results, with each point being an empirical estimate over $1000$ independent
experiments of the expected $\ell_2$-norm of the type error. The simulations were conducted for three 
privacy levels $\epsilon  = 0.1, 0.5$ and $1.0$, and over a wide range of sampling parameters 
$m$ at each level. The corresponding theoretical utility bounds (see Equation~(\ref{eqn:utilitybound}) of 
Theorem~\ref{thm:utility}) are illustrated by the solid curves, and the optimal number of samples 
(see Equation~(\ref{eqn:optimalsampling})) are shown with the vertical lines.

We make the following observations: Firstly, we observe that the 
theoretical prediction of the optimal number of samples aligns well with the experimental results.
In other words, the optimal sampling factor computed using the theoretical bounds is nearly identical
to that obtained via experiment, for all privacy levels.
Secondly, we find that the shape of the theoretical bounds is very similar to the shape formed by the experimental results, however the theoretical bounds are off by about a factor of $\sqrt{|\calX||\calY|}$.
To verify this, note that the shape of these bounds, when divided by a factor of $\sqrt{|\calX| |\calY|}$ and 
plotted with the dashed lines aligns well with the experimental results.
We confirmed that this behavior is reproduced even when we change the cardinality of the data. We observed this behavior over various cardinalities ranging from 12 to 768, with 1000 independent experiments conducted at each cardinality.

The looseness of the theoretical bounds can perhaps be explained by the bounding technique used in Equation~(\ref{eqn:NormBound}) on the ratio of $\ell_2$-norms,
\[
\frac{\| T_{\td X^m, \td Y^m} \|_2}{\| A T_{\td X^m, \td Y^m} \|_2}
\leq \frac{\sqrt{|\calX||\calY|} \| T_{\td X^m, \td Y^m} \|_1}{\| A T_{\td X^m, \td Y^m} \|_1}
= \sqrt{|\calX||\calY|},
\]
which introduces a pessimistic factor of $\sqrt{|\calX||\calY|}$ when bounding with the ratio of $\ell_1$-norms.
This pessimistic bound is approached only if $\| T_{\td X^m, \td Y^m} \|_2$ is close to $1$ (or, equivalently, $T_{\td X^m, \td Y^m}$ is close to a delta function) and $\| A T_{\td X^m, \td Y^m} \|_2$ is close to $1/\sqrt{|\calX||\calY|}$ (or, equivalently, $A T_{\td X^m, \td Y^m}$ is close to uniform).
Note that, while the gap can be made arbitrarily small, the bound cannot be met with exact equality due to the $\gamma$-diagonal structure of $A$ with $\gamma > 1$.
However, when the type of the data $T_{\td X^m, \td Y^m}$ is (or close to) uniform, the bound is loose as the ratio of $\ell_2$-norms is equal (or close) to one.
In our experiments, we have seen that the results with uniformly distributed synthetic data closely matches those with the real data, and appears to either match or bound the utility results for the other synthetic distributions.
If we tighten this bound by replacing the ratio of $\ell_2$-norms with one (assuming that this is a reasonable bounding approximation), the utility bound of Equation~(\ref{eqn:utilitybound}) becomes
\[
E \big\| \what T_{X^n,Y^n} - T_{X^n,Y^n} \big\|_2
  \leq \frac{c + 1}{\sqrt{m}},
\]
which reduces the overall error bound by roughly a factor of $\sqrt{|\calX||\calY|}$, since the condition number $c$ typically dominates over one.

\begin{figure}
\centering
\includegraphics[width=3.2in]{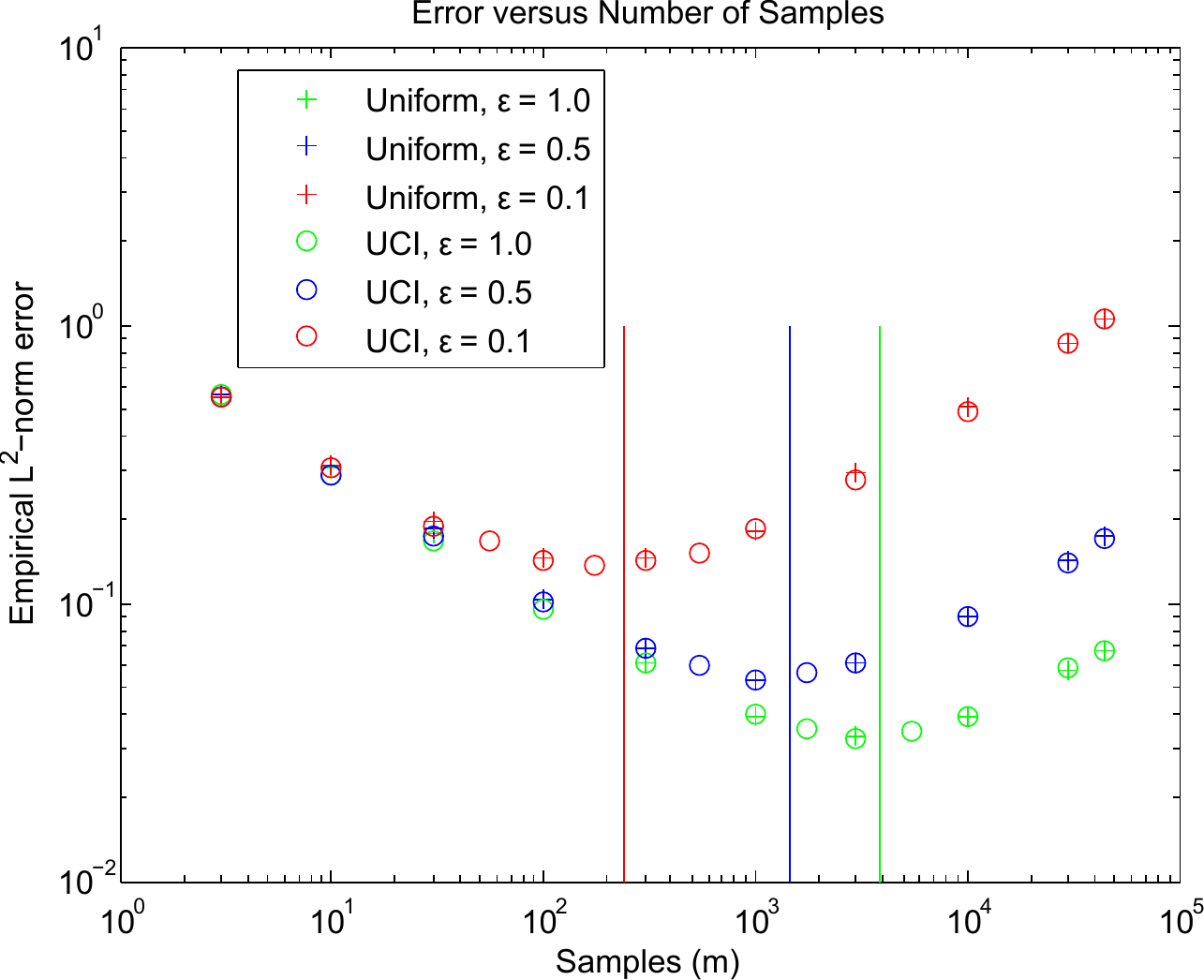}
\caption{The experimental results from simulations with the UCI dataset are plotted alongside the results from simulations with synthetic data with a ``uniform'' distribution. The vertical lines illustrate theoretically optimal number of samples at each privacy level. Each data point for both datasets was produced from $1000$ independent experiments.}
\label{fig:UCIUniformComp}
\end{figure}

\begin{figure}
\centering
\includegraphics[width=3.2in]{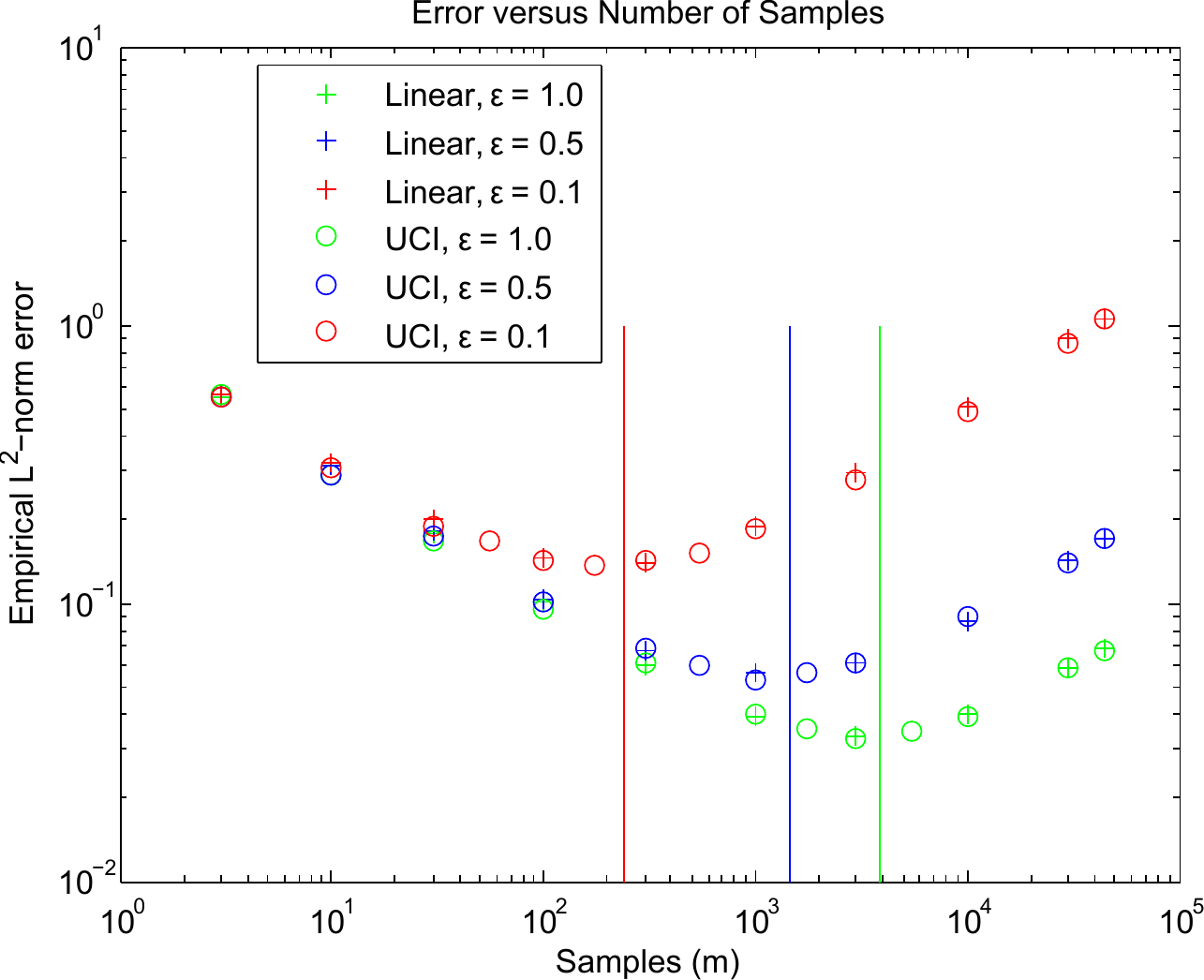}
\caption{The experimental results from simulations with the UCI dataset are plotted alongside the results from simulations with synthetic data with a ``linear'' distribution. The vertical lines illustrate theoretically optimal number of samples at each privacy level. Each data point for both datasets was produced from $1000$ independent experiments.}
\label{fig:UCILinearComp}
\end{figure}

\begin{figure}
\centering
\includegraphics[width=3.2in]{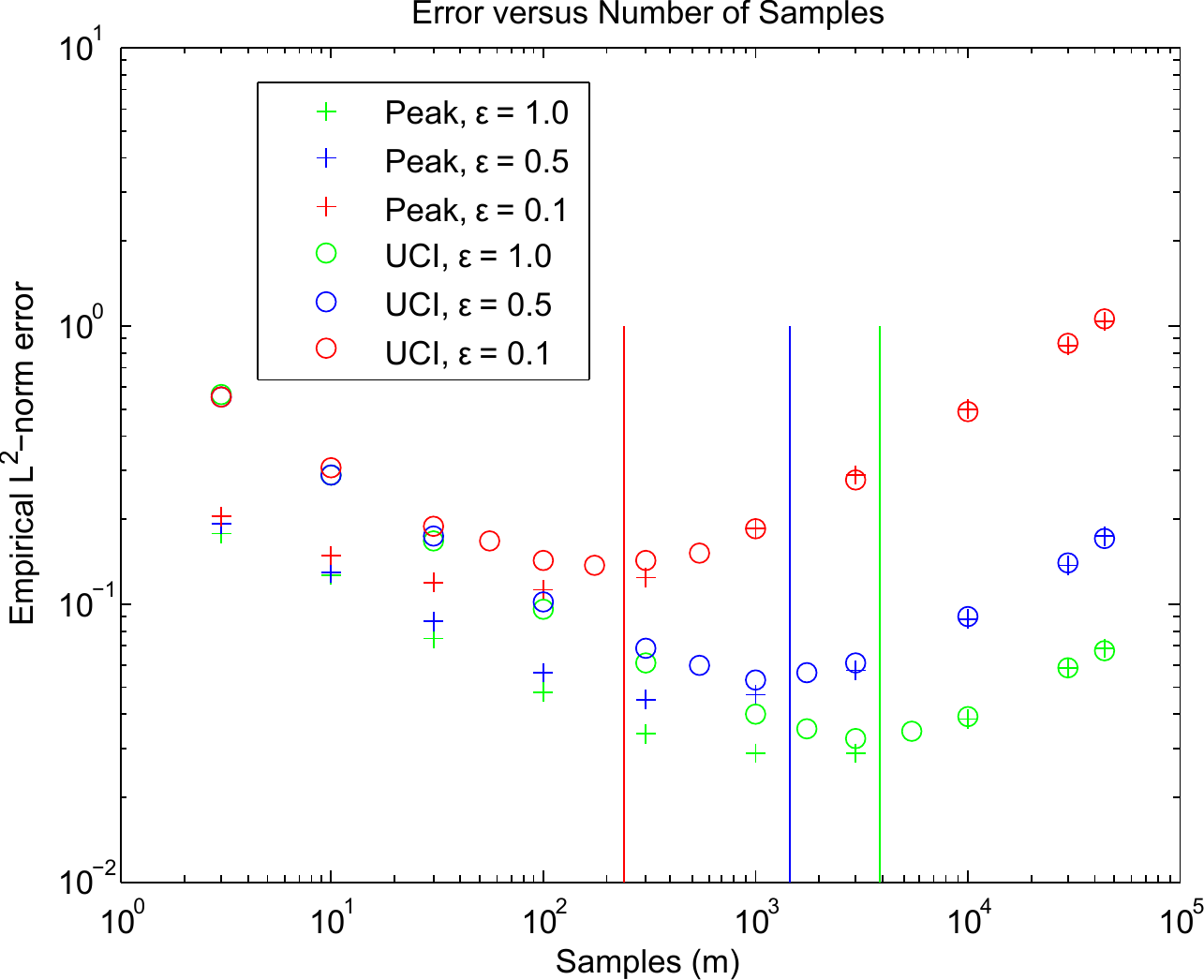}
\caption{The experimental results from simulations with the UCI dataset are plotted alongside the results from simulations with synthetic data with a ``peak'' distribution. The vertical lines illustrate theoretically optimal number of samples at each privacy level. Each data point for both datasets was produced from $1000$ independent experiments.}
\label{fig:UCIPeakComp}
\end{figure}

Next, we conducted simulations with synthetically generated data. We generated synthetic datasets of the same 
length as the UCI dataset ($n = 45222$) and cardinality ($|\calX||\calY| = 24$), but with three different distribution 
shapes, ``uniform'', ``linear'', and ``peaky''. The ``uniform'' dataset is simply uniformly distributed over the attribute set.
The ``linear'' dataset has a type function that linearly increases from $(1/q)$ for the least frequent attribute to $(24/q)$ for the most frequent attribute, where $q := (1 + \ldots + 24)$ is a normalizing constant. In the ``peaky'' dataset, the most
frequent attribute dominates the distribution at $90$ percent, while the other attributes uniformly share the remaining
$10$ percent of the distribution mass. The experiments with the ``uniform'' and ``linear'' synthetic datasets produced
results that were very similar to those with the UCI dataset. These results are plotted alongside the UCI dataset results
in Figures~\ref{fig:UCIUniformComp} and~\ref{fig:UCILinearComp}, respectively. However, the experiments with the
``peaky'' synthetic dataset, presented in Figure~\ref{fig:UCIPeakComp}, produced markedly different results than 
the UCI dataset experiments for lower values of $m$. We confirmed that this behavior is reproduced in 
experiments when the cardinality of the dataset is varied from 12 to 768. We conjecture that this is due to the 
high skewedness of the ``peaky'' synthetic dataset, which effectively reduces the impact of the cardinality of 
the dataset resulting in decreased error for lower values of $m$, the number of samples.

\begin{figure}
\centering
\includegraphics[width=3.2in]{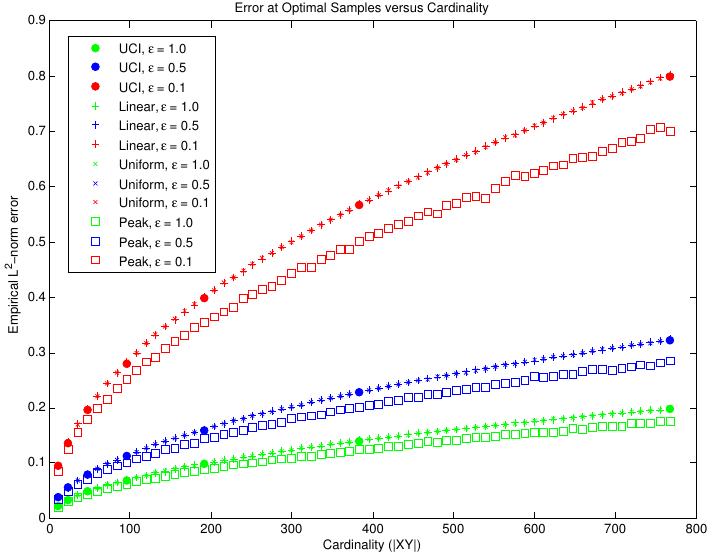}
\caption{The experimental results from simulations with the UCI dataset are plotted alongside the results from simulations with synthetic data. The number of samples for each pair of cardinality and privacy level is computed by Equation~(\ref{equ:optimalsamples}).  Each data point for both datasets was produced from $1000$ independent experiments.}
\label{fig:minimumerror}
\end{figure}

Our experiments confirm that using the optimal number of samples ($m^*$) derived from theoretical bound in 
Equation~(\ref{equ:optimalsamples}) consistently achieves near the minimum error in our experiments. This is observed
in all experiments with differential cardinalities, different data distributions and different privacy levels. We plot the 
$\ell_2$-norm error in the estimated joint distribution for the optimal number of samples $m^*$ in Figure~\ref{fig:minimumerror} for real and synthetic data experiments, at all three levels of privacy. The error curve of the UCI 
dataset overlaps with the error curve of the ``linear'' distribution and the error curve of the ``uniform'' distribution. 
The error of the ``peaky'' distribution is consistently lower than other distributions. As mentioned above, we 
conjecture that this is due to the high skewedness of this synthetic dataset which effectively reduced 
the impact of the cardinality on the utility measure.

\section{Discussion}
\label{sec:discussion}

We conclude our paper with a brief discussion to summarize our results and outline practical considerations toward implementing our proposed system.

\subsection{Summary of Results}

We analyzed a proposed system that combines sampling with PRAM to produce a privacy-preserving mechanism that enables data release for statistical analysis.
The sampling stage has two benefits in the system: 1) it enhances the system privacy improving the privacy-utility tradeoff, 2) it reduces the costs of one-time-pad encryption that provides strong security against a facilitating server.
Sampling reduces the amount of PRAM noise needed to provide a desired level of privacy, but oversampling will actually degrade the estimation performance since too much noise is required to maintain privacy.
However, undersampling will also degrade estimation performance since less data is gathered.
In this balance, there is an optimal sampling parameter, which we found in our analysis and confirmed in experiments with real and synthetic data.

\subsection{Practical Considerations}

The privacy-preserving framework described in this work is easy to implement in practice with
very small modifications to the abstract setting of this paper. For instance,
in the problem setting discussed above, encryption was accomplished by means of 
a one-time-pad which is an information-theoretic abstraction.
Actually using one-time-pads may be feasible if the sampling parameter is small enough to allow key distribution at a reasonable cost.
However, a practical alternative would be to perform encryption with a conventional stream cipher, with the key provided to the curators and the authorized
researcher but not to the server.
From the perspective of the authorized researcher and the
database respondents, the privacy-utility tradeoff remains the same. The only change is 
that, the data released by the curators has computational privacy instead of information theoretic
privacy against the server. In other words, a computationally bounded server cannot recover
the data sampled by the curators.

Furthermore, several interesting variants of the proposed framework are possible owing to the
fact that sampling, encryption and PRAM-based perturbation can commute without changing
the privacy-utility tradeoff.
The ordering of these operations is flexible allowing other architectures with the parties performing different roles.
For instance, if the curators want a secure external database storage facility, then they could encrypt the full database with a stream cipher, and request that the server perform both sampling and PRAM.

An important practical issue that has not been addressed in this work is the synchronization of the curators' databases and the sampling phase.
In our development, it is assumed that the respondents in Alice's and Bob's database are already synchronized and that they are able to sample in the same locations.
A practical approach toward database synchronization could involve using secure hashes of the unique IDs associated with each record, if available.
Synchronization of the sampling locations could be accomplished by either the curators directly sharing the sampling indices (using no more than $m \log n$ bits of communication) or by sharing the seed of a cryptographically secure pseudorandom number generator, that drives the choice of the sampling locations.
In the latter approach, the use of pseudorandomness would affect the statistical privacy guarantees against the researcher, however the practical impact would likely be insignificant against a computationally bounded researcher.
If the application allows for flexible architectures as described earlier, another alternative would be to have the sampling performed by the server.

\bibliographystyle{IEEEtran}
\bibliography{ref}

\end{document}